\documentclass[intlimits,twoside,a4paper]{article}

\usepackage{amsmath,amssymb}
\usepackage{graphicx}
\usepackage{wrapfig}

\usepackage[T2A]{fontenc}
\usepackage[cp1251]{inputenc}
\usepackage{cmpj2}

\issue{2013}{16}{1}{13201}
\doinumber{10.5488/CMP.16.13201}

\newcommand{\bs}{\boldsymbol}

\title{Nonequilibrium distribution functions of nucleons in relativistic
       nucleus-nucleus collisions}

\author{D. Anchishkin\refaddr{a1}, V. Naboka\refaddr{a2}, J. Cleymans\refaddr{a3}}

\authorcopyright{D. Anchishkin, V. Naboka, J. Cleymans, 2013}

\addresses{
\addr{a1}Bogolyubov Institute for Theoretical Physics, 03680 Kiev, Ukraine%
\addr{a2} Taras Shevchenko Kiev National University, 03022 Kiev, Ukraine
 \addr{a3}  University of Cape Town, Rondebosch 7701, South Africa }

\date{Received July 3, 2012, in final form July 31, 2012}

\begin{document}

\maketitle


\begin{abstract}
The collision smearing of the nucleon momenta about their initial
values during relativistic nucleus-nucleus collisions is investigated.
To a certain degree, our model belongs to the transport type, and we
investigate the evolution of the nucleon system created at a
nucleus-nucleus collision. However, we parameterize this development by
the number of collisions of every particle during evolution rather
than by the time variable.
It is assumed that the group of nucleons which leave the system after the same
number of collisions can be joined in a particular statistical ensemble.
The nucleon nonequilibrium distribution functions, which depend on
a certain number of collisions of a nucleon before freeze-out, are derived.


\pacs 25.75.-q, 25.75.Gz, 12.38.Mh
\end{abstract}

\section{Introduction}
\label{section:introduction}

The problem of isotropization and thermalization in the course of
collisions between heavy relativistic ions attracts much
attention, because the application of thermodynamic models is one
of the basic phenomenological approaches to the description of
experimental data.
Moreover, the assumption regarding a local thermodynamic equilibrium, along with
other factors, is successfully used in various domains of high-energy physics.
Meanwhile, many questions concerning this problem remain open
for discussion.

The main goal in the investigations of the collisions of relativistic nuclei
is to extract the pertinent physical information on the nuclear matter and its
constituents.
In the present paper we propose a transparent analytical model of
the nucleon spectrum which occurs in the course of relativistic heavy-ion collisions.
Our model is aimed at extracting the physical information from the
nucleon spectra which concerns the nonequilibrium processes.

Let us look at the cross-section of a multiparticle production during
the collision of two nuclei ``A'' and ``B'' (see figure~\ref{fig:two-nuclei}).
In order to describe of the nucleon subsystem one can parameterize the final
state of the nucleon ensemble by $4$-vector $K_N=(E_N,\, \bs K_N)$.
In the center of mass of this $N$-nucleon ensemble, where $\bs K_N=0$,
the total cross-section reads
\begin{eqnarray}
\left[ \frac{\rd^N \sigma_{\rm nucleon} }
{\rd^3p_1\, \rd^3p_2\, \ldots \rd^3p_{N}\, } \ \prod_{n=1}^N \omega(\bs p_n)\right]_{E_N}\,
=\, W(\bs p_1,\, \bs p_2, \ldots ,\, \bs p_{N})\,
\delta \left[ E_N - \sum_{n=1}^N \omega(\bs p_n) \right] \,,
\label{total-cs5}
\end{eqnarray}
where $\omega(\bs p)=\sqrt{m^2+\bs p^2}$ (in the final state, the particles are regarded as
free ones) and we adopt the system of units where the speed of light $c = 1$.
Due to the presence of the $\delta$-function, which ``fixes'' the energy of the
nucleon system, the last expression (\ref{total-cs5}) looks like a probability
in the microcanonical ensemble.
Then, it is reasonable to make the Laplace transform with respect to the energy
$E_N$ of the nucleon ensemble
\begin{eqnarray}
\left[  \frac{\rd^N \sigma_{\rm nucleon} }
{\rd^3p_1\, \rd^3p_2\, \ldots \rd^3p_{N}\, } \ \prod_{n=1}^N \omega(\bs p_n) \right]_\beta\,
=\, W(\bs p_1,\, \bs p_2, \ldots ,\, \bs p_{N})\, \prod_{n=1}^N \,
\re^{ -\beta \omega(\bs p_n) } \ .
\label{total-cs6}
\end{eqnarray}
It turns out that now one can describe the final state of the nucleon subsystem
through one of the two parameters: the total energy $E_N$ or the parameter
$\beta$.

All the above formulae were introduced for a brief discussion of the basics of the
statistical model (for details see \cite{lurcat-mazur-1964,hakim-2011}).
Actually, the statistical model neglects all the dynamics accumulated in
the transition probability $W$ in favor of the features of the phase space.
Formally this is expressed like approximation:
\begin{equation}
W(\bs p_1,\, \bs p_2, \ldots ,\, \bs p_{N}) \ \approx \ {\rm const} \,.
\label{sm-approx}
\end{equation}
Then, from (\ref{total-cs6}) one immediately obtains the multi-nucleon
cross-section parameterized by $\beta$:
\begin{equation}
\left[  \frac{\rd^N \sigma_{\rm nucleon} }
{\rd^3p_1\, \rd^3p_2\, \ldots \rd^3p_N } \ \prod_{n=1}^N \omega(\bs p_n) \right]_\beta
= \prod_{n=1}^N \, W_n \,
\re^{ -\beta \omega(\bs p_n) } \,,
\label{total-cs7}
\end{equation}
where $W_n$ are some constants and $\prod_{n=1}^N  W_n=\mathrm{const}$.

On the other hand, the statistical description, which arises after freeze-out,
is conceptually restricted just to several conserved quantities: total energy of
the system $E_N$, number of particles $N$, and conserved charges such as the
baryon number.
Of course, this dependence can be transferred to the descriptions by means of
conjugate quantities: parameter $\beta$ and chemical potentials which are in
correspondence with $N$ and the conserved charges.
As we see from (\ref{total-cs7}), this description provides a certain information
on the spectrum of the registrated particles.
Meanwhile, any dynamical information on multiscattering processes during
collisions is lost.
At the same time, it is well understood that a microscopic description can be
carried out just on some level of approximation.
For instance, if it is possible to factorize the transition probability $W$,
i.e., to write it in the form
$ W(\bs p_1,\, \bs p_2, \ldots ,\, \bs p_{N}) \approx \prod_{n=1}^N \, W(\bs p_n) $\,,
we come to factorization of the multi-nucleon cross-section for the particles of one species
\begin{equation}
\left[  \frac{d^N \sigma_{\rm nucleon} }
{\rd^3p_1\, \rd^3p_2\, \ldots \rd^3p_{N}\, } \prod_{n=1}^N \omega(\bs p_n) \right]_\beta
= \ \prod_{n=1}^N \, \left[ \, W(\bs p_n) \re^{ -\beta \omega(\bs p_n) }  \,
\right] \,.
\label{total-cs8}
\end{equation}
The approximation of the sequential rescatterings of a particle during collision
of nuclei which is proposed in the present paper
is exactly in this framework. We follow the chain of reactions (rescatterings)
of every nucleon that goes through a number of hadron transformations, and we obtain a single-particle spectrum of the nucleon $W_{_M}(\bs p)$ which
depends on the number $M$ of collisions (reactions) of the nucleon (of the baryon
characterized by the baryon number $B=1$).

We argue that the number of nucleon collisions (reactions) at AGS and SPS
energies is finite and the maximal number of collisions $M_{\rm max}$ is not
more than $M_{\rm max}=43$.
Apart from this, all the nucleons which are freezed out during a particular
nucleus-nucleus collision, can be subdivided into groups. In every group, the
nucleons went through the same number $M$ of collisions.
We determined that the most po\-pu\-la\-ted groups are for the number of collisions $M$
which are in the range: $M=14\, -\, 17$.
Starting from the initial moment of the nucleus-nucleus collision, we follow the
sequential collisions of every nucleon through the opposite nucleus (see
figure~\ref{fig:two-nuclei}). Nevertheless, the original nucleon can be transformed
during a particular collision into another particle, for instance into delta isobar
$\Delta^+$.
Then, we follow a new particle which carries the same charges (the baryon number,
electrical charge, etc.) as the original nucleon.
During the last collision (it can be a decay), all these ``intermediate''
particles transform back into nucleons. Hence, we can investigate just nucleons in the
final state.
Starting from this point, every group of nucleons in the final noninteracting state is the
subject of a statistical model.
For the multiscattering stage of evolution, we treat UrQMD  \cite{urqmd1,urqmd2} as a
source of ``experimental data'' which we use to adjust the parameters of our model.

We propose a mutually complementary combination of these two approaches, i.e.,
an approximate description of the dynamical stage of evolution of
nucleons during the nucleus-nucleus collision which is completed with a
statistical description of the nucleon freezed out stage.
Our approach is based on ``The multiscattering-statistical model'' elaborated by us.

\section{The multiscattering-statistical model}
\label{section:model}

\begin{figure}[h!]
\vspace{-20mm}
\includegraphics[width=0.47\textwidth]{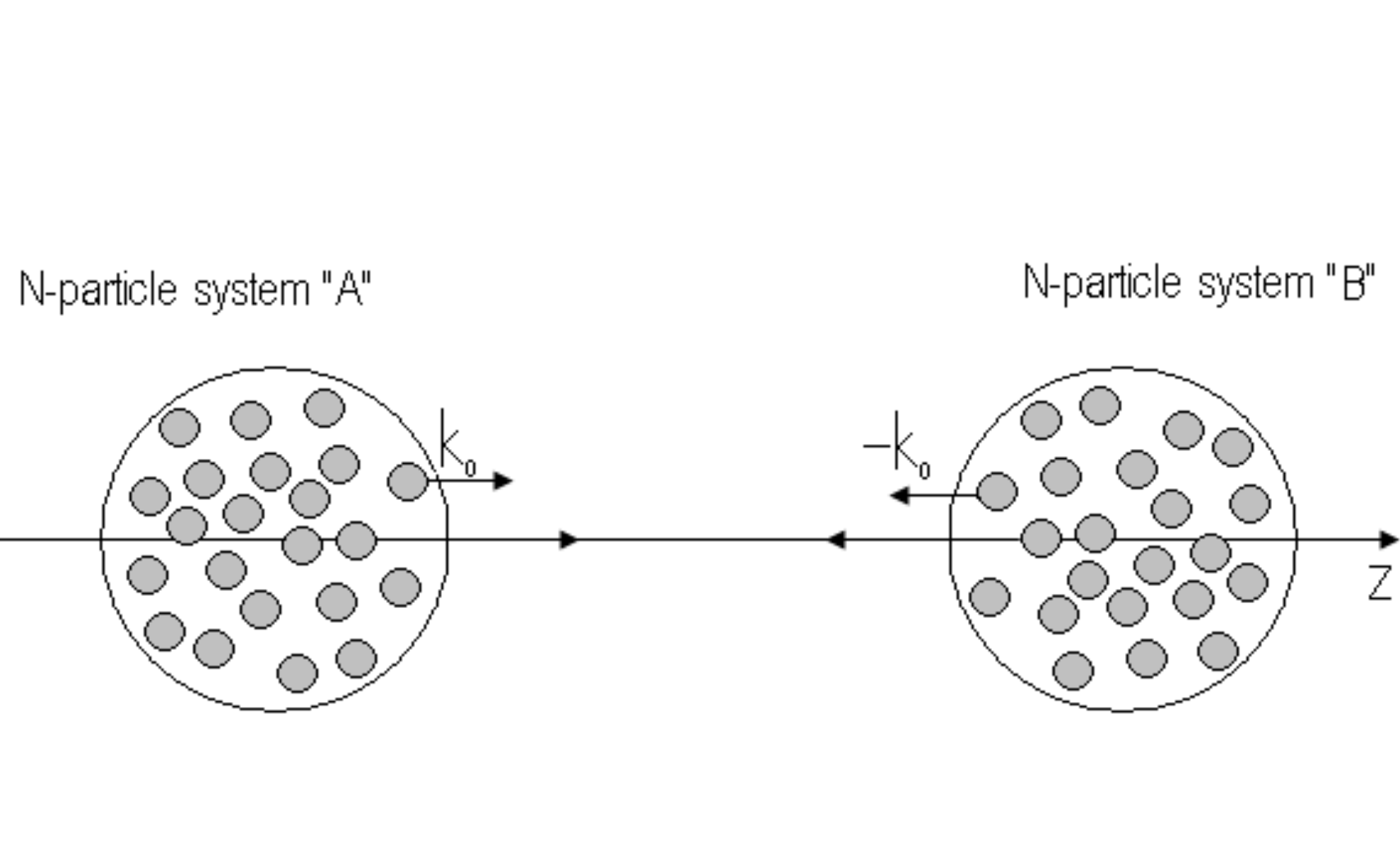}
\hfill
\includegraphics[width=0.5\textwidth,height=65mm]{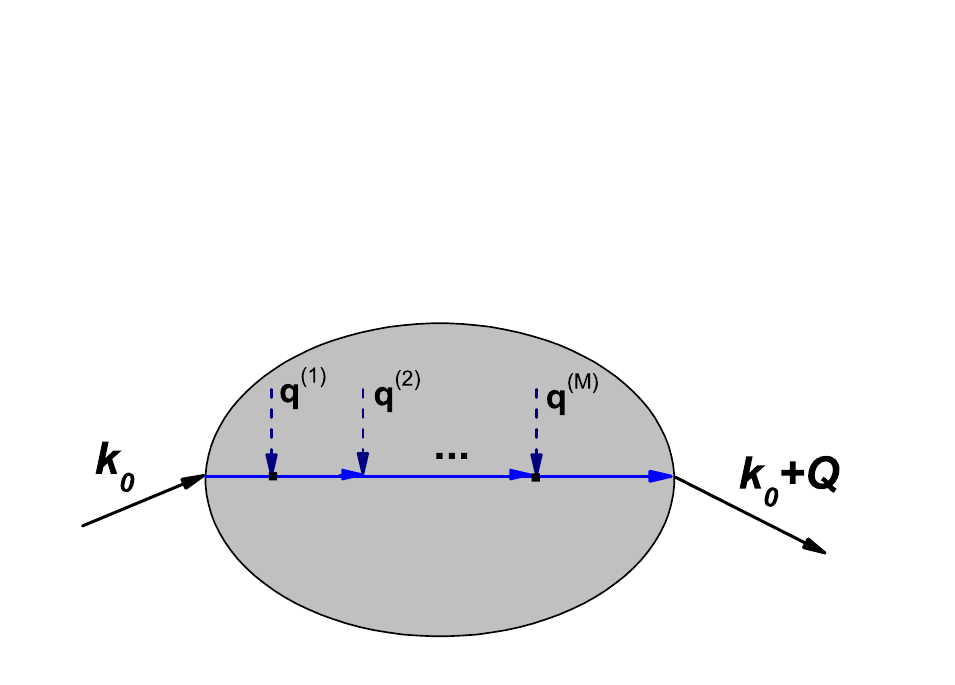}
\\
\parbox[t]{0.47\textwidth}{\caption {\small
Two colliding identical nuclei.
Two-stream system is created during the collision of every nucleon from nucleus
``A'' with nucleons from nucleus ``B'' and vice versa.}
\label{fig:two-nuclei} }
\hfill
\parbox[t]{0.5\textwidth}
{\caption{\small
Transformation of the initial nucleon momentum, $\bs k_0$, as a result of $M$
collisions; ${\bs Q}=\sum_{m=1}^M{\bs q}_m$ is the total momentum transfer
after $M$ collisions, $\bs q_m$ is the momentum transfer in the $m$-th
collision. }
\label{fig:totalmt}  }
\end{figure}

Consider successive variations of the momentum of a nucleon from nucleus $A$
(see figure~\ref{fig:two-nuclei})
which moves along the collision axis from left to right through the nucleus $B$.
Every $m$-th collision induces the momentum transfer, $\bs q_m$, for this
nucleon.
Consequently, after $M$ collisions, the nucleon acquires the momentum~$\bs k$:
\begin{equation}
{\bs k}_0 \quad \rightarrow \quad {\bs k}_0+{\bs q}_1
\quad \rightarrow  \quad
{\bs k}_0+{\bs q}_1 + {\bs q}_2
\quad \rightarrow \quad \cdots  \quad \rightarrow
\quad {\bs k}_0 + \bs Q \, =\, \bs k \,.
\label{fin-momentum}
\end{equation}
where $\bs Q=\sum_{m=1}^M \bs q_m$ is the total momentum transfer finally
obtained by our nucleon after $M$ collisions (see figure~\ref{fig:totalmt}).
If the $M$-th collision is the last one, then the nucleon is free after having
been freezed out from the system. As a matter of fact, it will be a group of such nucleons which experienced the same number of collisions $M$ before the freeze-out.
The relation of these nucleon groups (nucleon sub-ensembles) to the spectrum
is discussed in the next section.

%

We assume that all the momentum transfer ${\bs q}_m$ obtained by the nucleon from the nucleus ``A''
during its travel through the system ``B'' are some random quantities.
Then, when the number of collisions $M$ is big enough
in accordance with the central limit theorem, the distribution of the
random quantity $\bs Q$ should be a Gaussian one
\begin{equation}
\mathbb{G}(\bs Q) = \frac1z \exp{\left\{- \frac{\big(\bs Q - \left\langle \bs Q
\right\rangle \big)^2}{2\sigma^2} \right\}} \!
\quad \rightarrow \quad \!
\mathbb{G}_M(\bs k) = \frac 1{z_{_M}} \exp{\left\{- \frac{\big(\bs k -
\bs k_0-\left\langle \bs Q \right\rangle_M \big)^2 }{2\sigma_M^2} \right\}}  ,
\label{gauss-distrib}
\end{equation}
where we take into account the equation~(\ref{fin-momentum}), i.e., $\bs Q=\bs k - \bs k_0$,
and  explicitly write the dependence on the final momentum of the nucleon $\bs k$.
Here, $z$ is the normalization constant.
Actually, this distribution depends on the number $M$ of random
quantities $\bs q_1,\, \bs q_2,\, \ldots ,\, \bs q_M$, which coincides with the number of collisions $M$  experienced by the nucleon before being freezed out.

Further, we design the many-particle distribution function $\mathbb{G}_M(E_M; {\bs k}_1, {\bs k}_2, \ldots, {\bs k}_N)$ as
\begin{equation}
\mathbb{G}_M(E_M;\, \widetilde{\bs k}) \
=\ \frac 1{A_M(E_M)}\, \widetilde{\mathbb{G}}_M(E_M;\, \widetilde{\bs k})  \,,
\qquad \widetilde{\bs k}=(\bs k_1,\, \bs k_2,\, \ldots ,\, \bs k_N),
\label{gauss-distrib4}
\end{equation}
where
\begin{equation}
\widetilde{\mathbb{G}}_M(E_M;\, \widetilde{\bs k}) \,
\equiv \,
\prod_{n=1}^N \exp{\left\{- \frac{\big(\bs k_n - \bs k_0-
\left\langle \, \bs Q\, \right\rangle_M \big)^2}{2\sigma_M^2} \right\}} \,
\delta\left[ E_M - \sum_{j=1}^N \omega(\bs k_j) \right] \,,
\label{gauss-distrib4a}
\end{equation}
and make the Laplace transform with respect to the total energy $E_M$
\begin{equation}
\widetilde{\mathbb{G}}_M(\beta; \widetilde{\bs k})
= \! \int \! \rd E_M \, \re^{-\beta\, E_M}\, \widetilde{\mathbb{G}}_M(E_M; \widetilde{\bs k}),
\qquad
Z_M(\beta) = \! \int \! \rd E_M \, \re^{-\beta\, E_M} A_M(E_M) .
\label{l-transform2}
\end{equation}
Basically, from now on, any physical quantity that depends on the nucleon
momenta, can be averaged using the many-particle distribution function
\begin{equation}
\mathbb{G}_M(\beta;\, \widetilde{\bs k}) \ \equiv \
\frac 1{Z_M(\beta)} \,
 \widetilde{\mathbb{G}}_M(\beta;\, \widetilde{\bs k})\,.
\label{l-transform3}
\end{equation}

If we introduce expression (\ref{gauss-distrib4a}) into (\ref{l-transform2}),
then in the framework of the multiscattering-statistical model (MSS-model) we obtain
\begin{equation}
\mathbb{G}_M(\beta_{_M};\, \widetilde{\bs k})
=
\frac 1{Z_{_M}(\beta_{_M})}\, \prod_{n=1}^{N_{_M}}
\exp{\left\{\, - \beta_{_M} \omega(\bs k_n)- \frac{\big(\bs k_n -
\bs k_0-\left\langle \, \bs Q\, \right\rangle_M \big)^2}{2\sigma_M^2} \right\}}
= \prod_{n=1}^{N_{_M}} \, f_M(\bs k_n) ,
\label{mp-df}
\end{equation}
where $Z_{_M}(\beta_{_M})=\left[z_{_M}(\beta_{_M})\right]^{N_{_M}}$ and the
single-particle distribution function is as follows:
\begin{equation}
f_M(\bs k) \ \equiv \
\frac 1{z_{_M}(\beta_{_M})} \, \exp{\left\{\, - \beta_{_M} \omega(\bs k)-
\frac{\big(\bs k - \bs k_0-\left\langle \, \bs Q\, \right\rangle_M \big)^2}
{2\sigma_M^2}\, \right\}}
\label{sp-df}
\end{equation}
with $z_{_M}(\beta_{_M})$ as the single-particle partition function.
Note, a derivation of the analogous distribution can be found in~\cite{anchishkin-muskeyev-yezhov-2009-1,anchishkin-muskeyev-yezhov-2009-2}.


\section{Two-stream dynamics}
\label{section:sub-ensembles}

Based on the  obtained results, we are coming to a description of a two-stream
system which is created in the course of relativistic nucleus-nucleus collisions.
The description is based on the following assumptions:
\begin{enumerate}
  \item Starting from the initial state (first touch of the colliding nuclei),
  at an arbitrary moment of time, there are two systems of nucleons: one system
  consists of nucleons with a positive
        $z$-component of the nucleon momentum, i.e., $p_z \geqslant 0$ (we refer to this
        system as ``A'') and the second system consists of the nucleons with
        a negative $z$-component of the nucleon momentum, i.e. $p_z \leqslant 0$ (we
        refer to this system as ``B'', see figure~\ref{fig:two-nuclei}).
        Even after the freeze-out, there is a good enough separation of these systems
        along $p_z$-axis.
  \item The number of collisions of every nucleon (hadron) is finite because the
        lifetime of the fireball is limited.
        To determine the maximal number of collisions, $M_{\rm max}$, in a
        particular experiment we use the results of UrQMD simulations
        \cite{urqmd1,urqmd2}.
  \item Since the colliding nuclei are  spatially restricted,  different
        nucleons experience a different number of collisions, and
        it is intuitively clear that the collision histories of the inner nucleons
        and surface nucleons will be different.
        That is why, we subdivide all the freezed out nucleons (nucleon
        ensemble) into different nucleon sub-ensembles in accordance with the number of
        collisions $M$ the nucleons went through before being freezed out.
        Then, the nucleons from every sub-ensemble give their own contribution
        to the total nucleon spectrum.
\end{enumerate}

It is time to write a two-stream distribution function $\mathbb{F}_M(\bs p)$
which, in accordance with the first assumption, should take into account the
spectrum produced from both particle streams, ``A'' and ``B''.
Being normalized to unity, a two-stream distribution function looks as follows:
\begin{equation}
\mathbb{F}_M(\bs p) \
=\  \frac12 \, \left[\, f_M^{(a)}(\bs p) \, +\, f_M^{(b)}(\bs p) \, \right] \,,
\label{ts-df}  
\end{equation}
where
\begin{equation}
f_M^{(a)}(\bs k)  =
\frac 1{z_{_M}(\beta_{_M})} \,
\exp{\left\{ - \beta_{_M} \omega(\bs k)-
\frac{\bs k_\perp^2}{2\left( \sigma^2_\perp \right)_M}\, \right\}} \,
\exp{ \left\{ - \frac{\big(k_z - k_{0z} - \left\langle Q_z \right\rangle_M \big)^2}
{2\left( \sigma^2_z \right)_M}\, \right\} }
\label{ts-df4}  
\end{equation}
and
\begin{equation}
f_M^{(b)}(\bs k)  =
\frac 1{z_{_M}(\beta_{_M})} \,
\exp{\left\{ - \beta_{_M} \omega(\bs k)-
\frac{\bs k_\perp^2}{2\left( \sigma^2_\perp \right)_M}\, \right\}} \,
\exp{ \left\{
- \frac{\big(k_z + k_{0z} + \left\langle Q_z \right\rangle_M \big)^2}
{2\left( \sigma^2_z \right)_M}\, \right\} }
\label{ts-df5}
\end{equation}
with $\bs k_0 = \big(0,\, 0,\, k_{0z}\big)$ and we assume $
\langle \sigma^2_x \rangle_M \, =\, \langle \sigma^2_y \rangle_M \,
\equiv \, \langle \sigma^2_\perp \rangle_M $\,.
Here, the single-particle partition function reads
$ z_{_M}(\beta_{_M}) = \int \rd^3k/(2\pi)^3 f_M^{(a)}(\bs k) $.
    Now, the slope parameter $\beta_{_M}$ reflects also a collective motion of the
    $M$-th nucleon sub-ensemble moving in the laboratory system.
%
\begin{figure}[h!]
\includegraphics[width=0.48\textwidth]{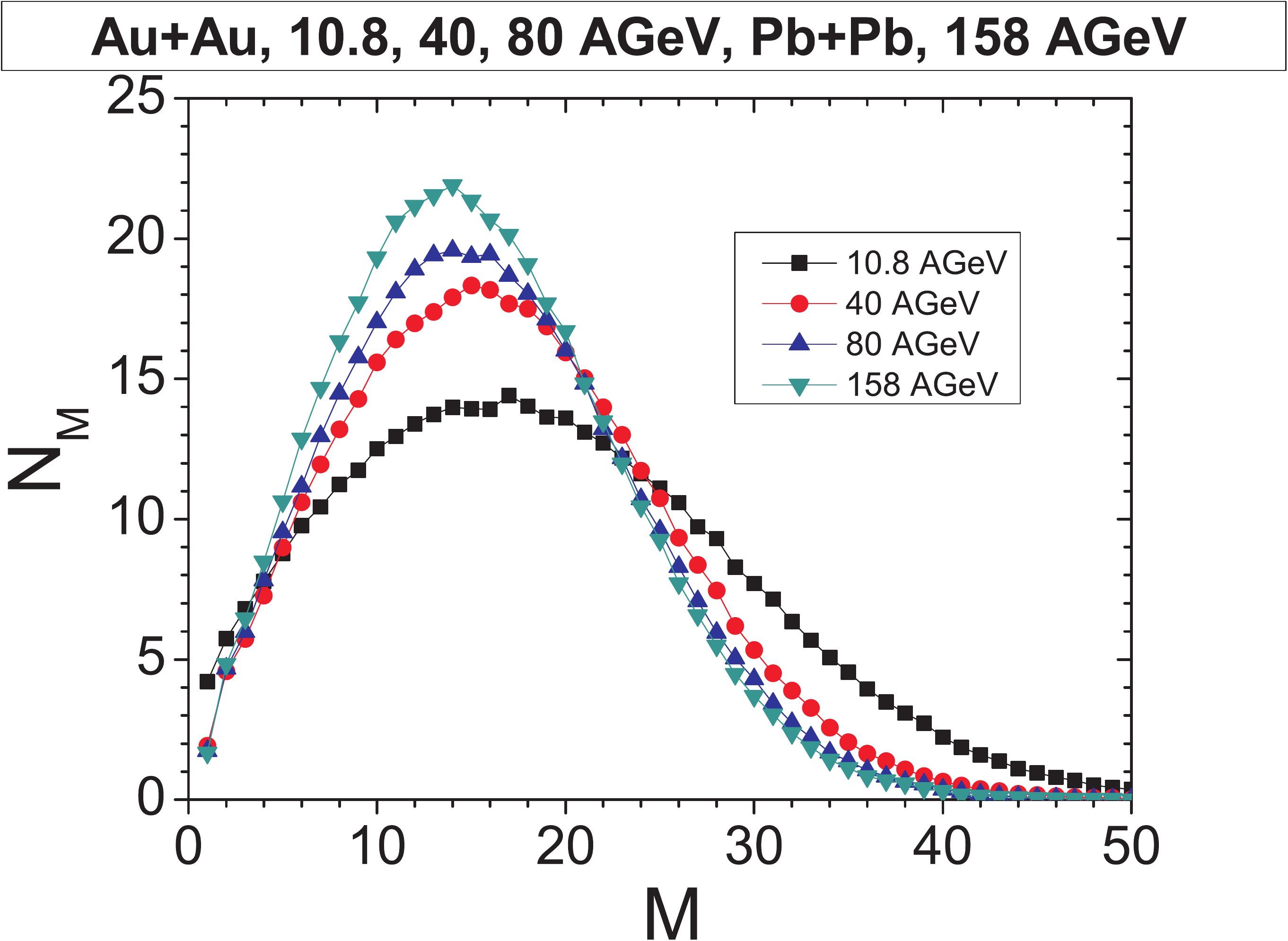}
\hfill
\includegraphics[width=0.44\textwidth]{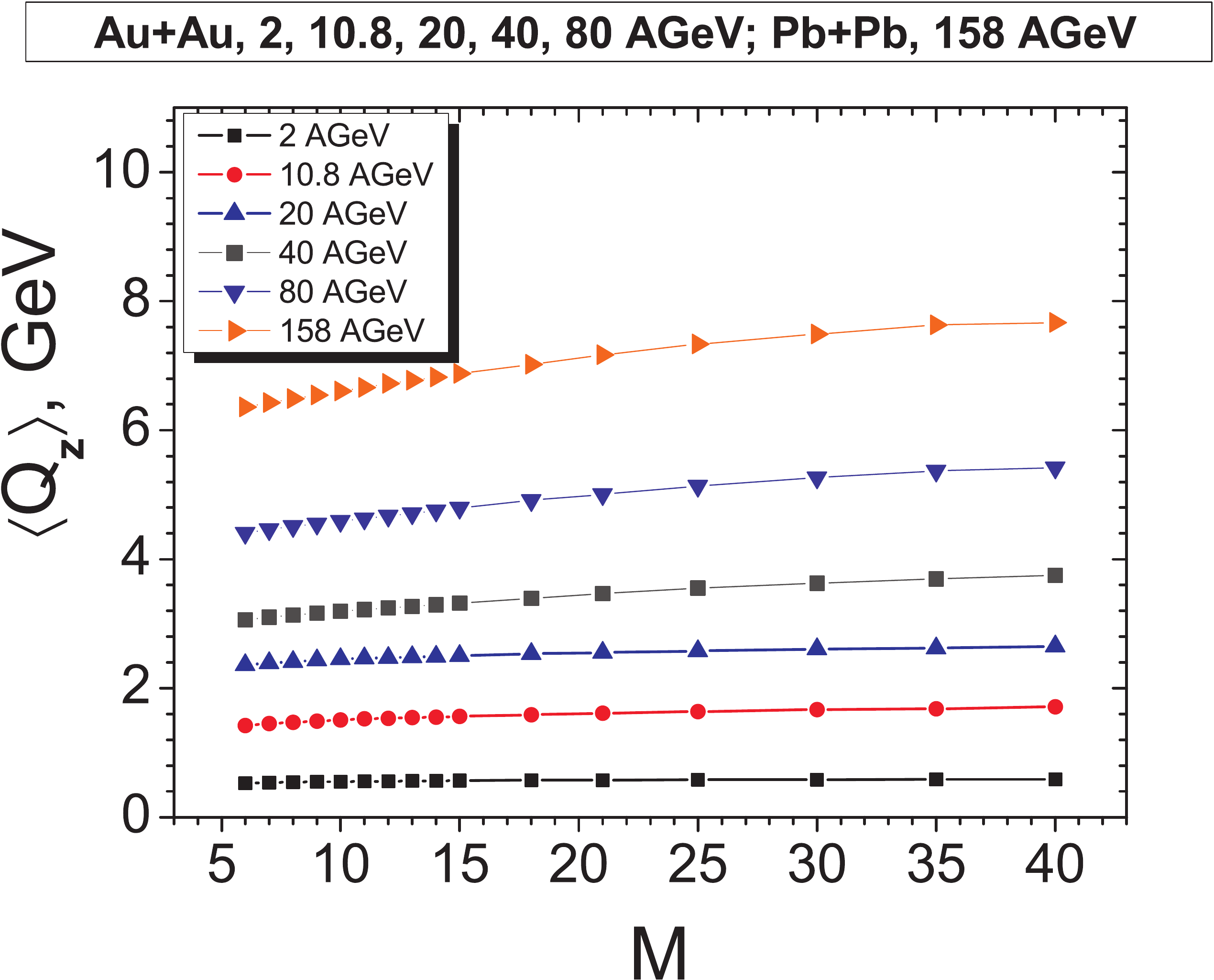}
\\
\parbox[t]{0.48\textwidth}{
\caption{\small (Color online)
The populations $N_M$ of the nucleon sub-ensembles which depend
on the number of collisions $M$.
The result is obtained from the UrQMD simulations
for the most central collisions. }
\label{fig:populations}   }
\hfill
\parbox[t]{0.48\textwidth}{
\caption{\small (Color online)
The $z$-component of the mean momentum transfer versus the number of collisions.}
\label{fig:qz_eff} }
\end{figure}
%

Keeping in hands the two-stream distribution functions $\mathbb{F}_M(\bs p)$,
where $M=1,\,2,\, \ldots \,M_{\rm max}$, one can construct the nucleon spectrum
which occurs in the course of a central nucleus-nucleus collision.
If we denote the number of particles in a particular sub-ensemble, where the
nucleons experienced $M$ collisions, by $N_M$, then in the c.m.s. of the colliding
nuclei, the total nucleon spectrum is as follows:
\begin{equation}
\frac{\rd N}{\rd^{\,3} p} \ = \ \sum_{M=1}^{M_{\rm max}}\, N_M \, \mathbb{F}_M(\bs p)\,,
\quad {\rm with} \quad
\sum_{M=1}^{M_{\rm max}} \, N_M \ = N_{\rm total} \,,
\label{spectrum}
\end{equation}
where $N_{\rm total}$ is the total number of net nucleons.
The number of nucleons $N_M$ in every sub-ensemble calculated for different
energies with the help of the microscopic transport model UrQMD
\cite{urqmd1,urqmd2} is depicted in figure~\ref{fig:populations}.
Every sub-ensemble of nucleons can be described as an ideal gas which moves with
some collective velocity.

\section{Extraction of physical parameters from the data}
\label{section:extraction}

\begin{figure}[h!]
     \includegraphics[width=0.52\textwidth]{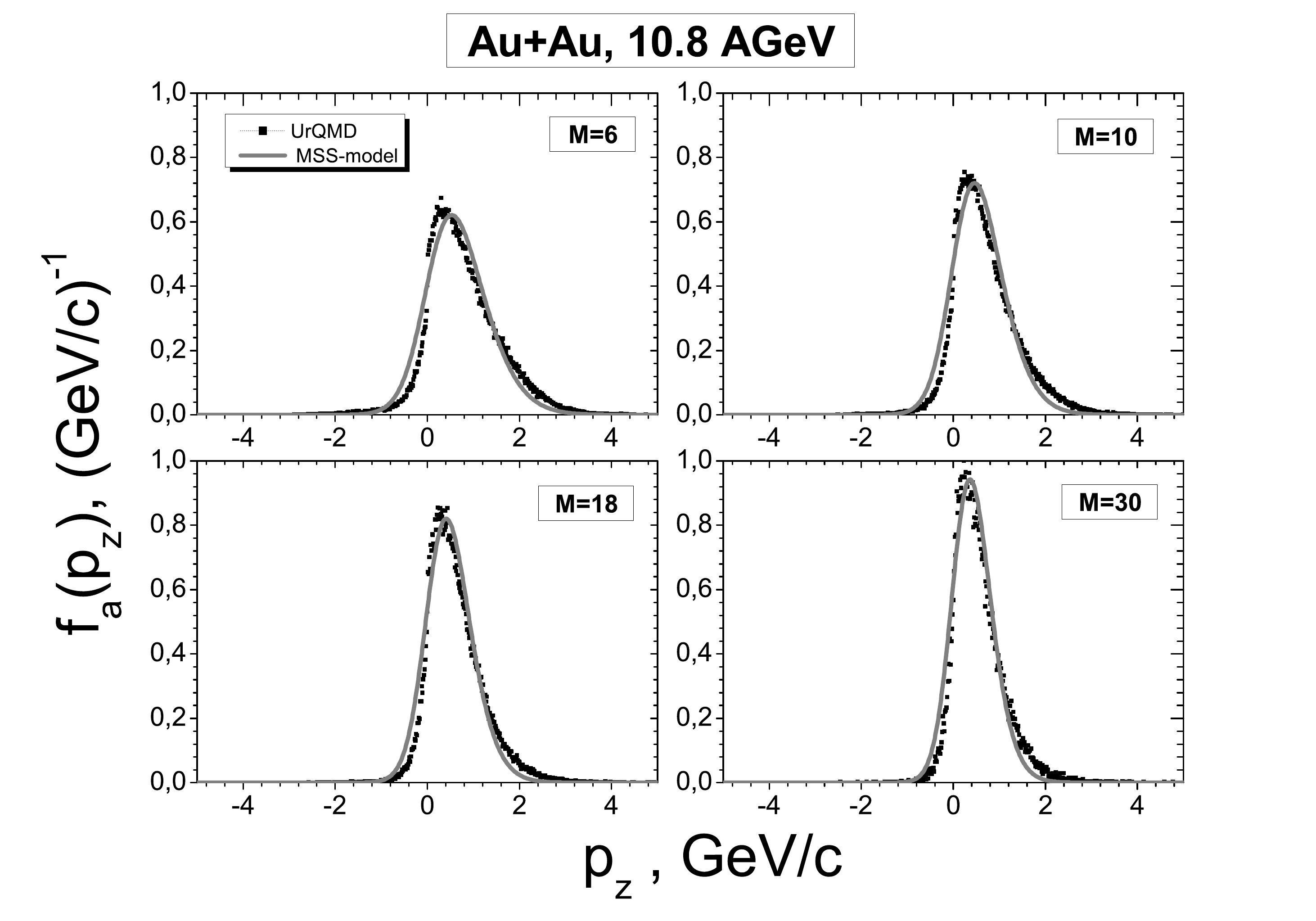}
     \includegraphics[width=0.52\textwidth]{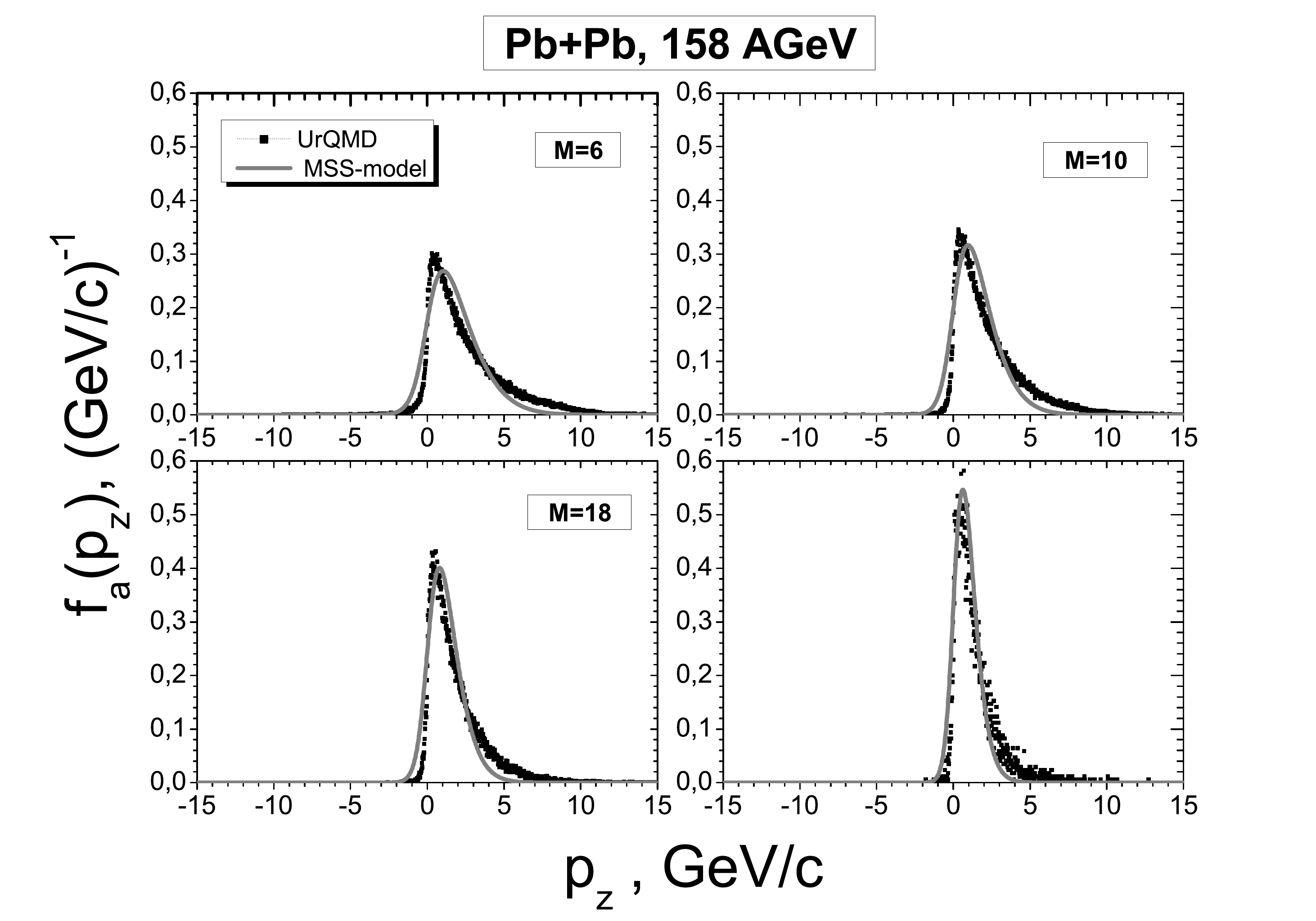}
     \caption{\small
The spectrum of the $M$-th  nucleon sub-ensemble ($M=6,\, 10,\, 18,\, 30$) with
respect to $p_z$-component of the nucleon momentum calculated using the UrQMD
transport model for Au+Au collision (black squares).
The grey curves are the fits of the UrQMD data within the framework of the proposed
multiscattering-statistical model (MMS-model). }
\label{fig:fit2}
\end{figure}

First we obtain from UrQMD \cite{urqmd1,urqmd2} the longitudinal distribution
of nucleons for every $M$-th sub-ensemble.
The distributions for the stream ``A'' (positive $p_z$) and the
stream ``B'' (negative $p_z$) were obtained separately. We refer to this pool of
distributions as ``UrQMD data''.
We fit the ``UrQMD data'' on the longitudinal distribution
  of nucleons of the   $M$-th sub-ensemble of the stream ``A'' exploiting the
  theoretical distribution function (\ref{ts-df4}) integrated over a transverse momentum.
  The variations of the theoretical distribution function were provided by
  four parameters:
  $\left\langle Q_z \right\rangle_M$, $\beta_{_M}$, $\left(\sigma^2_\perp \right)_M$
  and $\left(\sigma^2_z \right)_M$.
The results of the fit of the ``UrQMD data'' (nucleon longitudinal distributions)
for the energies $10.8$ and $158$~AGeV are depicted in figure~\ref{fig:fit2}.
%
%

\begin{figure}[h!]
     \includegraphics[width=0.52\textwidth]{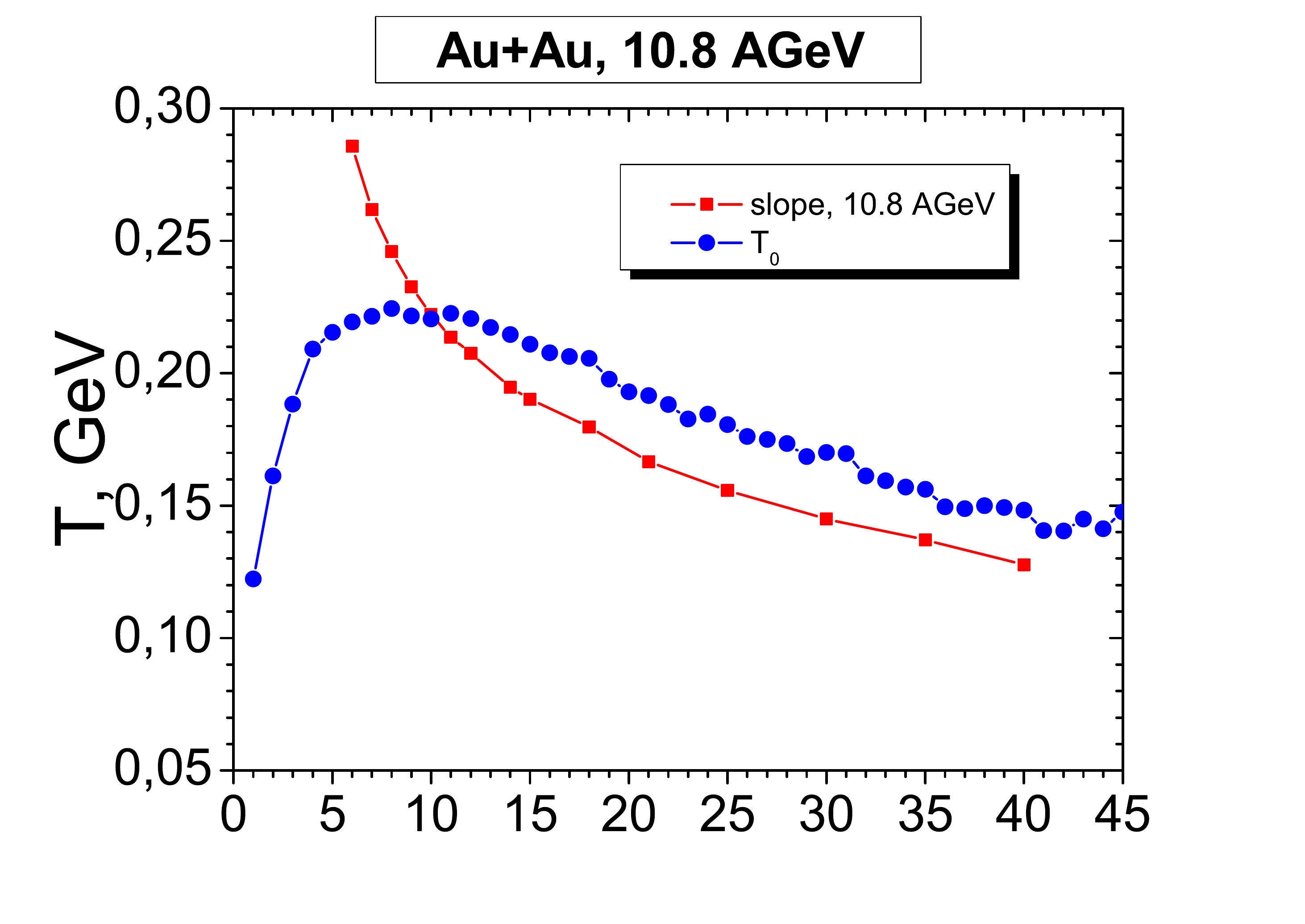}
     \includegraphics[width=0.52\textwidth]{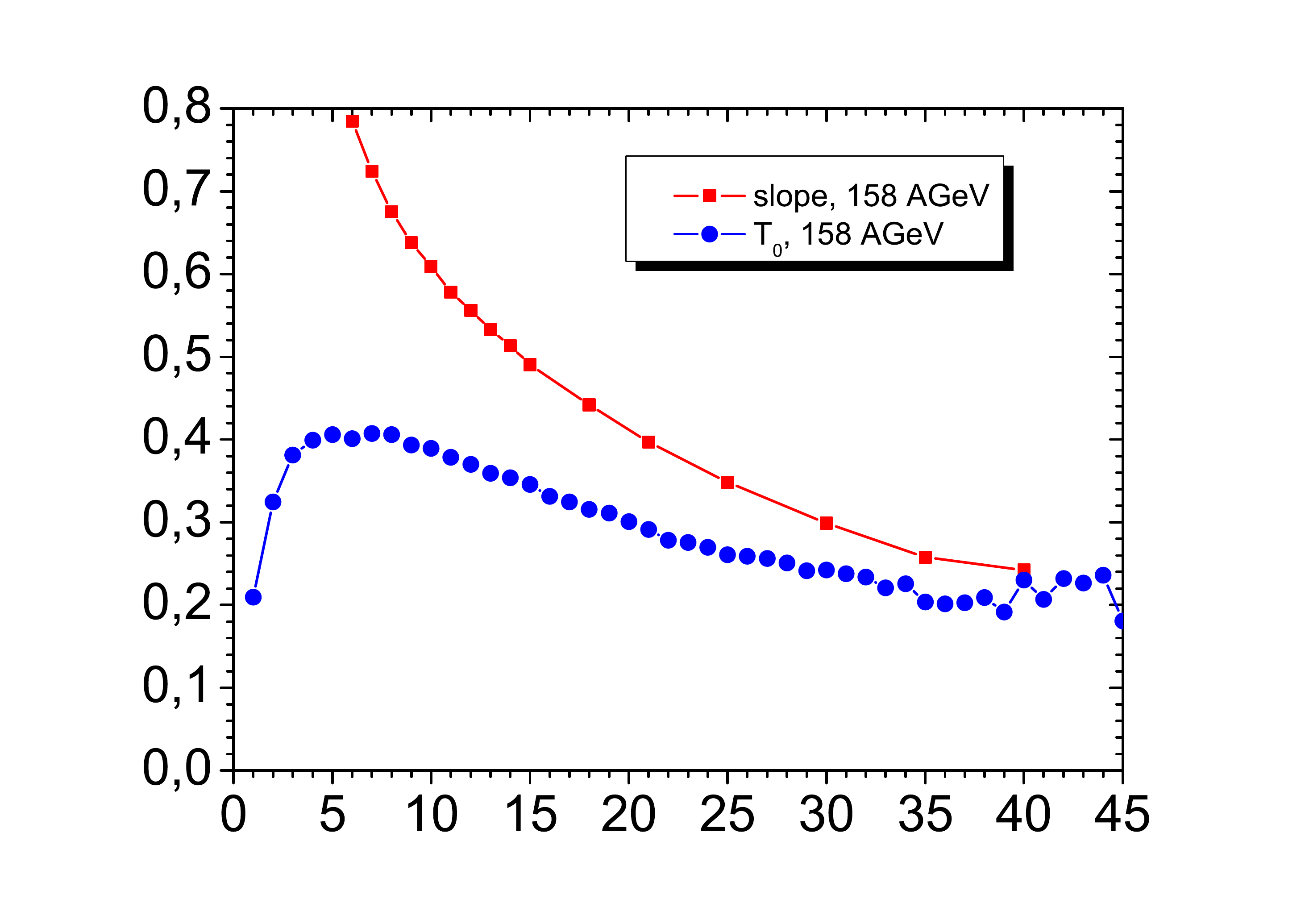}
     \caption{(Color online) Dependence of the slope parameter  $T_{_M}$ (curve marked with squares) and
     temperature $T_0$ (curve marked with circles)
     on the collision number $M$ for different energies of the
     nucleus-nucleus collision: 10.8, 158 AGeV. }
\label{fig:slope-parameter}
\end{figure}

The dependence of the parameters with respect to the number of reactions, $M$,
of the nucleon before the freeze-out is shown in the following figures:
The slope parameter $T_{_M}=1/\beta_{_M}$ in figure~\ref{fig:slope-parameter}
(the curves marked with square symbols)
and the temperature $T_0$ (the curves marked with circle symbols);
The mean value of the shift of the distribution function,
$\left\langle Q_z \right\rangle_M$, in figure~\ref{fig:qz_eff};
The longitudinal variance $\left(\sigma^2_z \right)_M$ in figure~\ref{fig:all-sigma}.
Note, we use the system of units where the Boltzmann constant is unit,
$k_{\rm B} = 1$.

Temperature of the hot ideal gas $T_0$ is determined in the local rest frame of the gas. This
temperature is connected with the total kinetic energy of $N$  particles
(nucleons) in the local rest frame in the following way:
%
\begin{equation}
\frac{E_{\rm r.f.}}{N} \ = \
3T_0 \, +\, m_{_N}\, K_1\left(\frac{m_{_N}}{T_0}\right)\Big/
K_2\left(\frac{m_{_N}}{T_0}\right) \,.
\label{determ-temper}
\end{equation}
In our model $E_{\rm r.f.}$ is the total kinetic energy of the sub-ensemble
of nucleons in the rest frame of this group of particles, $m_{_N}$ is the
nucleon mass ($E_{\rm r.f.} = \gamma \big[ E - V \, P_z \big] \,, \
\gamma = 1/\sqrt{1-V^2}$).
%
%

\begin{figure}[h!]
     \includegraphics[width=0.52\textwidth]{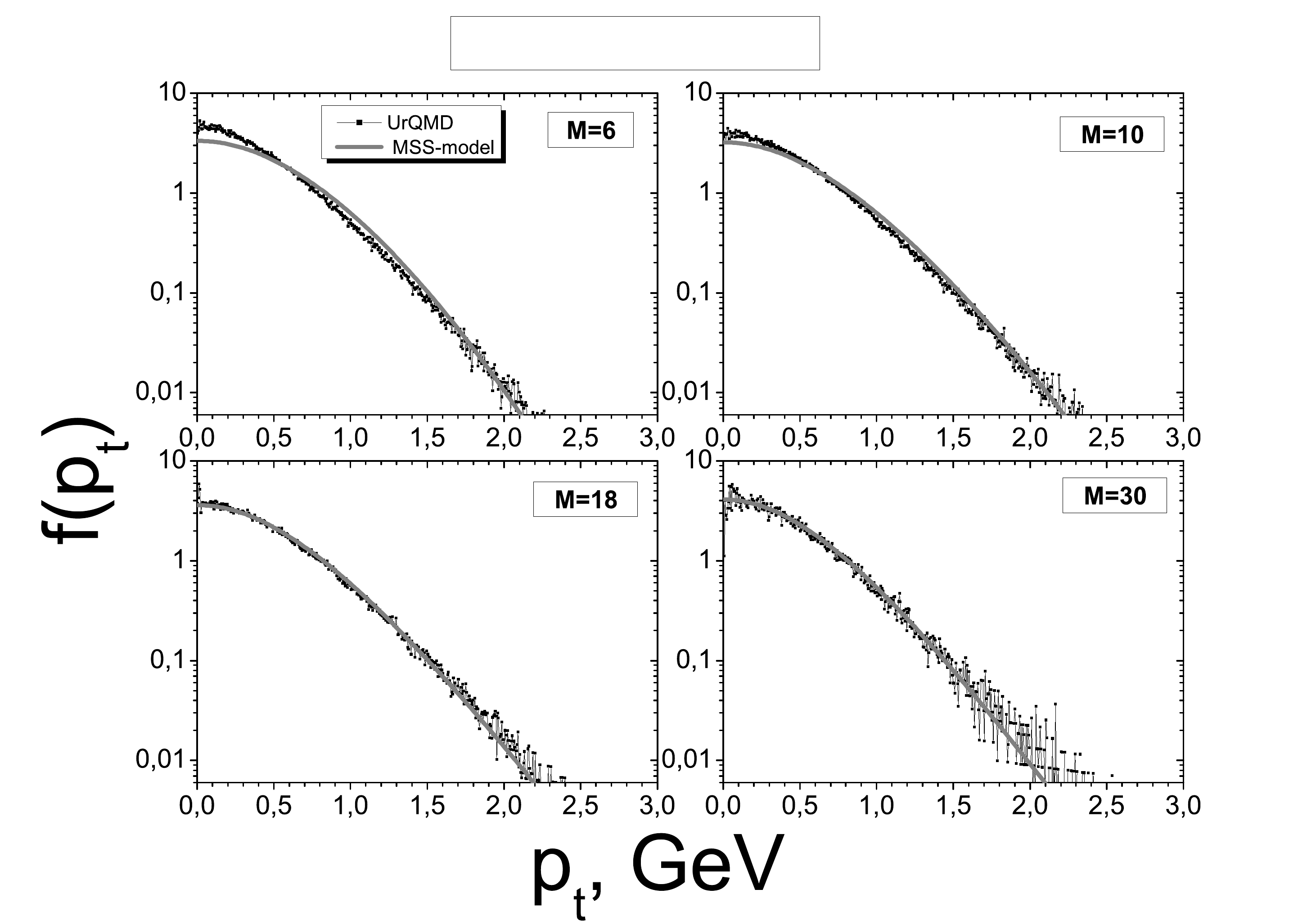}
     \includegraphics[width=0.52\textwidth]{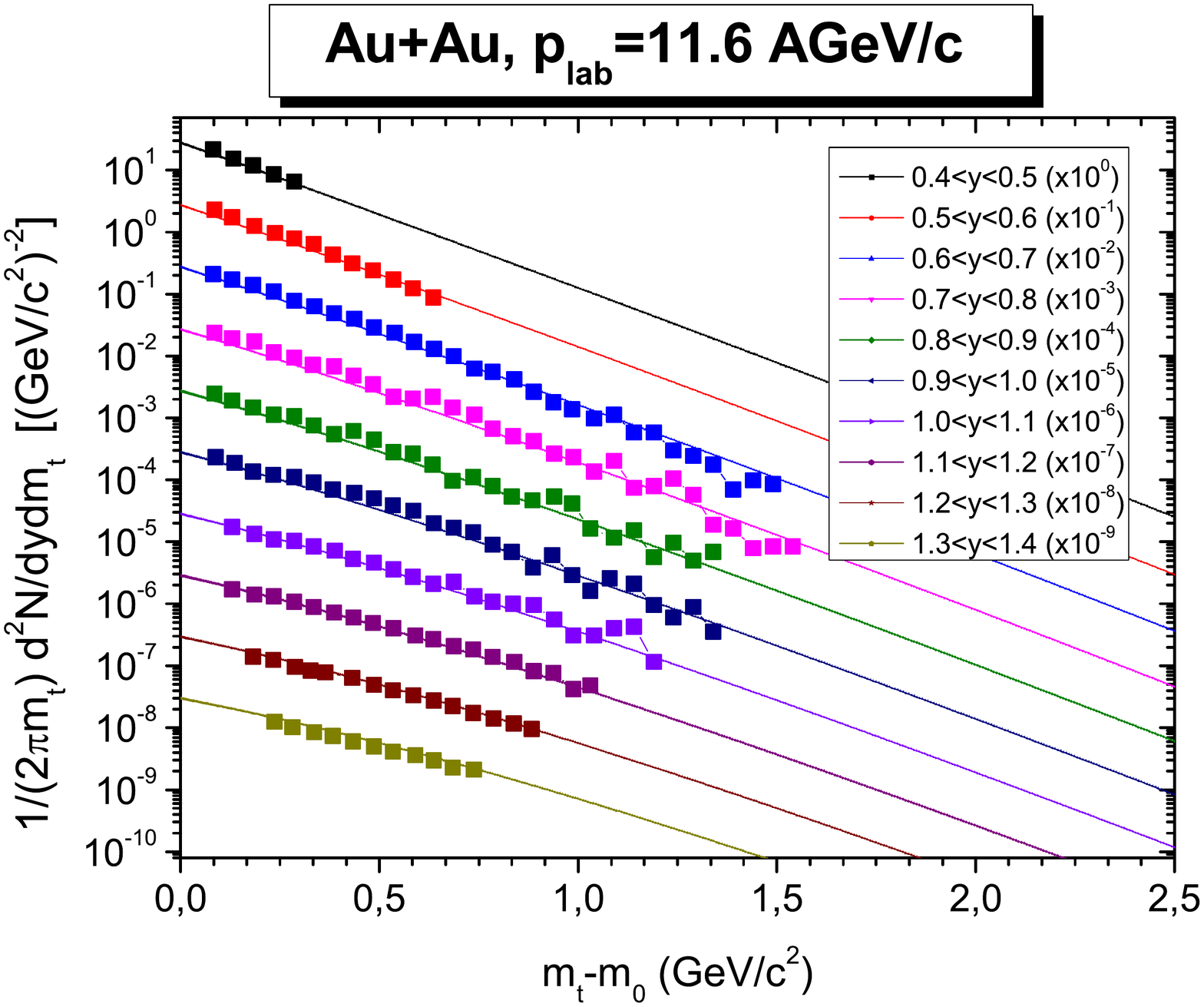}
          \caption{(Color online) Fit of the UrQMD data (left panel) and description of the experimental
     data \cite{E-802} on nucleon      transverse distribution in central Au+Au
     reactions at $p_{\rm lab}=11.6$A GeV/c (right panel).  }
\label{fig:exp-data}
\end{figure}

To obtain the transverse distribution  we integrate the ``A''-stream distribution function
(\ref{ts-df4}) over the longitudinal component of the nucleon momentum.
The results of the description of the ``UrQMD data'' on the nucleon transverse
distribution for the energy $20$~AGeV is depicted in
figure~\ref{fig:exp-data} (left hand panel).
For this description, we use the same values of parameters
$\left\langle Q_z \right\rangle_M$,
$\beta_{_M}$, $\left(\sigma^2_z \right)_M$ and $\left(\sigma^2_\perp \right)_M$,
which were obtained during the fit to the ``UrQMD data'' on nucleon longitudinal
distributions.
%
%
The experimental data for transverse nucleon distributions in central Au+Au
reactions at $p_{\rm lab}=11.6$A GeV/c \cite{E-802} were described using the formula (\ref{spectrum}), the result is depicted in
figure~\ref{fig:exp-data} (right hand panel).
We see a good agreement of the description with experiment.


\section{Discussion and conclusions}
\label{section:discussion}

The description of a many-particle system, which is in thermal equilibrium
state, can be regarded as quite understandable and complete by means of the temperature and
chemical potential if the latter is needed.
Then, to obtain the value of the temperature, one has to fit the particle spectrum
data using one of single-particle distribution functions.
The fitting procedure is nothing more as an extraction of the physical quantity, i.e.,
``temperature'', from the data.
\begin{figure}[h!]
     \includegraphics[width=0.5\textwidth]{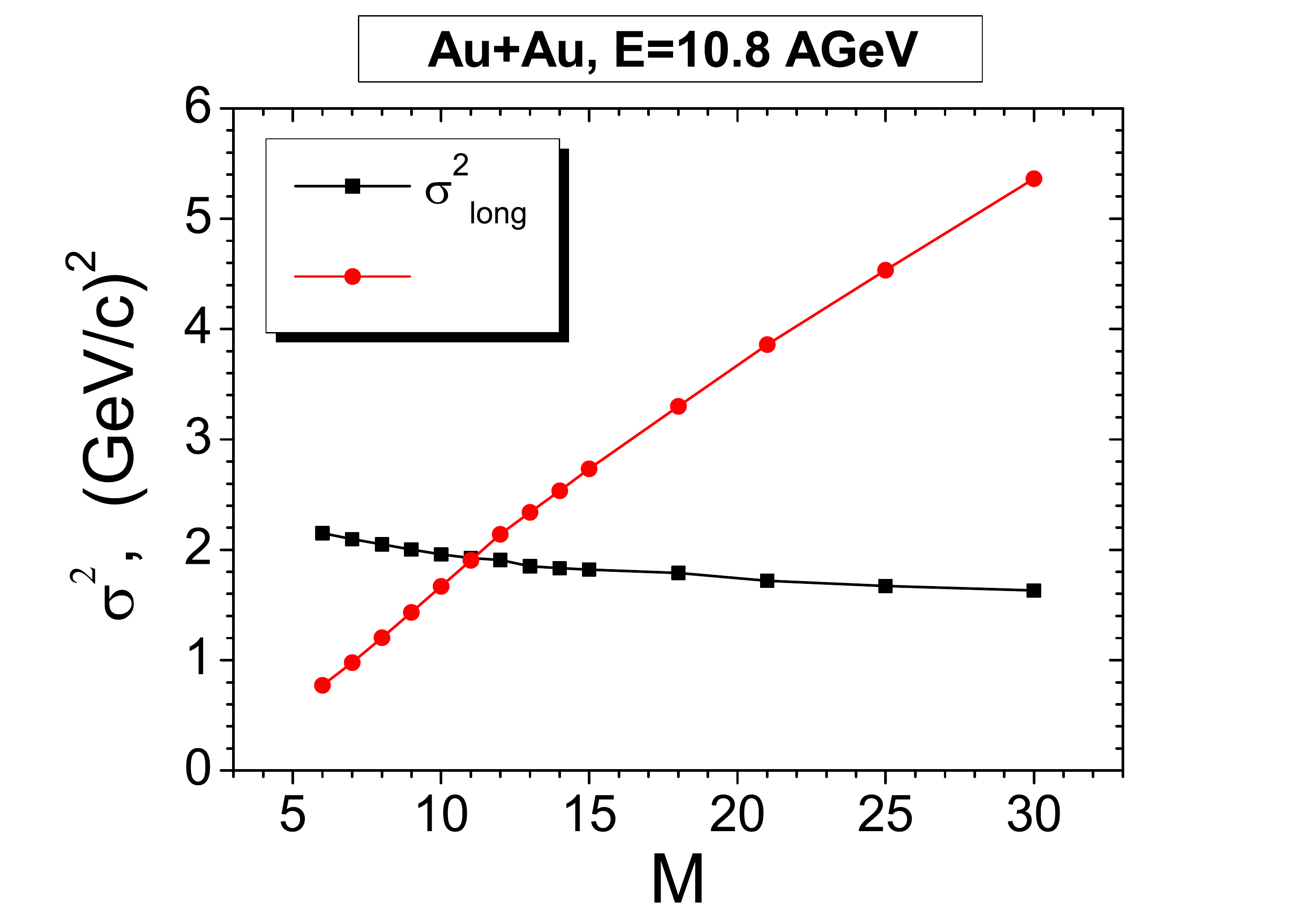}
     \includegraphics[width=0.5\textwidth]{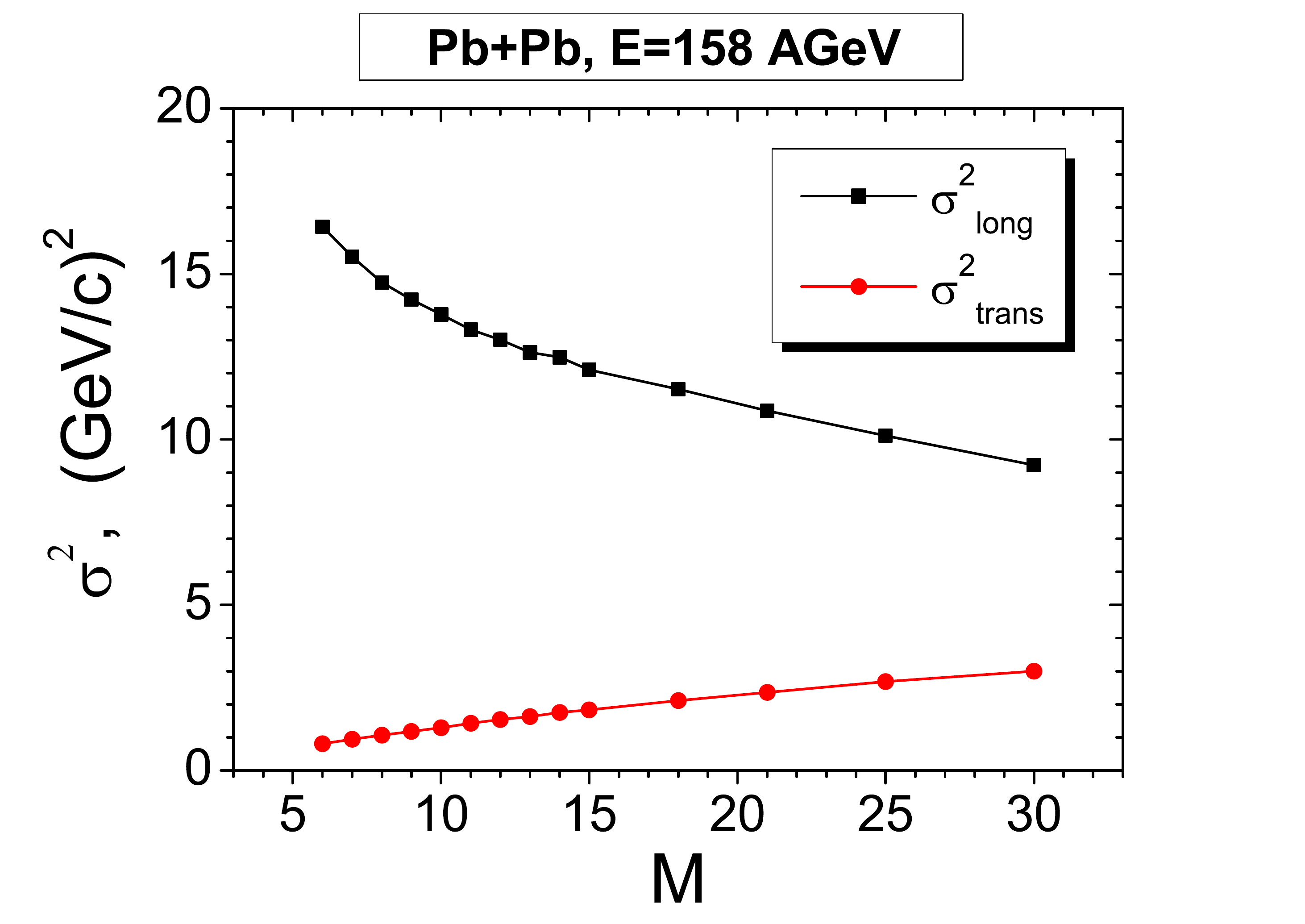}
     \caption{\small (Color online)
Longitudinal and transverse variances,
$\left(\sigma^2_z \right)_M$ and $\left(\sigma^2_\perp \right)_M$,
versus the number collisions $M$ for collision energies
$10.8$ and $158$~AGeV.  }
\label{fig:all-sigma}
\end{figure}

Next,  if we come to the description of a many-particle system in a
nonequilibrium (nonstationary, nonhomogeneous) state, a natural question
arises: which set of parameters
is needed to get a relevant physical picture of the many-particle system, which state evolves in time? Of course, by the words
``physical picture'' one means a physical
interpretation of the parameters of a model.
The set of parameters of a model, as well as the behavior of the evolution of the parameters
take the form of a specific language based on which we investigate, describe and ``understand''
our nonequilibrium many-particle system and the processes inside it.

In the present paper for description of a nonequilibrium state we propose three parameters, which
    are defined in ``The Multiscattering-Statistical  Model''.
\begin{enumerate}
  \item The slope parameter $T(M) = 1/\beta_M$, which reflects as well a collective motion of the nucleon
       sub-ensemble. Its dependence on the number of collisions $M$ experienced by nucleons, which
belong to $M$-th sub-ensemble, is depicted in figures~\ref{fig:slope-parameter}.
By means of the energy per particle (after the freeze-out), this parameter is
directly connected with the instant temperature $T_0(M)$ of the $M$-th nucleon sub-ensemble
which is defined in the frame where the sub-ensemble is in rest.
  \item The mean value of $z$-component of the total momentum transfer
        $\left\langle Q_z \right\rangle_M=\sum_{m=1}^M \langle q_z \rangle_m$, which
is related to the kinematics of a particle-particle collision.
  \item \looseness=-1The variance of the Gaussian distribution.
Actually, due to the different collision dynamics along the different axis, the variance is
split in two pieces:
a) The longitudinal variance $\left(\sigma^2_z \right)_M$.
b) The transverse variance $\left(\sigma^2_\perp \right)_M$.
In figures~\ref{fig:all-sigma}, the behavior of the transverse variance is
confronted with the longitudinal variance.
We see a steady broadening of the transverse distribution with an increase of the
number of collisions $M$ in the nucleon sub-ensemble.
Indeed, the initial transverse variance of nucleons is approximately zero
on the scale of collision energy.
The distribution of nucleons around the initial value in the transverse
direction becomes broader and broader with time.
\end{enumerate}

We can conclude that the evolution of the physical parameters of
``The Multiscattering-Statistical Model'' elaborated in
the present paper gives a transparent insight into the dynamics of the net
nucleons in the course of relativistic nucleus-nucleus collisions.
The appearance of the Gaussian distribution as a factor in the nonequilibrium
distribution function of nucleons, see (\ref{ts-df4}), is common
in describing any multiscattering process with a big but finite number of rescatterings
of the particle when we can regard every particular scattering independent of
others.
This condition is  especially satisfied for the nucleons in the course of high energy heavy-ion
collisions when the particle wavelength $\lambda=\hbar/p$ is much smaller than
the mean distance between the nucleons in a nucleus.

\section*{Acknowledgements}
\noindent
D.A. was supported by the program \textquotedblleft
Microscopical and phenomenological models of fundamental physical processes at
micro and macro scales\textquotedblright\
(Department of Physics and Astronomy of the NAS of Ukraine).


\newpage

\ukrainianpart
\vspace{-5mm}
\title{Нерівноважні функції розподілу нуклонів при релятивістичних ядро-ядерних зіткненнях}

\author{Д. Анчишкін\refaddr{a1}, В. Набока\refaddr{a2},  Ж. Клейманс\refaddr{a3}}

\addresses{
\addr{a1} Інститут теоретичної фізики ім. М.М. Боголюбова,
03680 Київ, Україна
\addr{a2} Київський національний університет імені Тараса Шевченка,
      03022 Київ, Україна
\addr{a3} Університет Кейп Тауна, Рондебош 7701, Південно-Африканська
      Республіка}

\makeukrtitle

\begin{abstract}
\tolerance=3000%
 Досліджується розмиття імпульсів нуклонів навколо своїх початкових
 значень, яке відбувається в релятивістичних ядро-ядерних зіткненнях.
 Наша модель відноситься, певною мірою, до транспортних, ми
 дослідили еволюцію нуклонної системи, створеної в
 ядро-ядерних зіткненнях, але ми параметризуємо цей розвиток не часом,
 а числом зіткнень кожної частинки.
 Припускається, що група нуклонів, які залишають систему зазнавши
 однакову кількість зіткнень, можуть бути об'єднані в певний статистичний
 ансамбль.
 Обраховується нерівноважна функція розподілу нуклонів в імпульсному
 просторі, яка залежить від певного числа зіткнень нуклона перед
 випромінюванням із системи.
\keywords релятивістичні зіткнення, нерівноважна функція розподілу,
 спектр нуклонів, параметризація еволюції

\end{abstract}

\lastpage

\end{document}